\PassOptionsToPackage{unicode}{hyperref}
\PassOptionsToPackage{hyphens}{url}
\PassOptionsToPackage{dvipsnames,svgnames,x11names}{xcolor}
\documentclass[
]{article}
\usepackage{xcolor}
\usepackage[margin=1in]{geometry}
\usepackage{amsmath,amssymb}
\usepackage{iftex}
\ifPDFTeX
  \usepackage[T1]{fontenc}
  \usepackage[utf8]{inputenc}
  \usepackage{textcomp} % provide euro and other symbols
\else % if luatex or xetex
  \usepackage{unicode-math} % this also loads fontspec
  \defaultfontfeatures{Scale=MatchLowercase}
  \defaultfontfeatures[\rmfamily]{Ligatures=TeX,Scale=1}
\fi
\usepackage{lmodern}
\ifPDFTeX\else
\fi
\IfFileExists{upquote.sty}{\usepackage{upquote}}{}
\IfFileExists{microtype.sty}{% use microtype if available
  \usepackage[]{microtype}
  \UseMicrotypeSet[protrusion]{basicmath} % disable protrusion for tt fonts
}{}
\makeatletter
\@ifundefined{KOMAClassName}{% if non-KOMA class
  \IfFileExists{parskip.sty}{%
    \usepackage{parskip}
  }{% else
    \setlength{\parindent}{0pt}
    \setlength{\parskip}{6pt plus 2pt minus 1pt}}
}{% if KOMA class
  \KOMAoptions{parskip=half}}
\makeatother
\usepackage{graphicx}
\makeatletter
\newsavebox\pandoc@box
\newcommand*\pandocbounded[1]{% scales image to fit in text height/width
  \sbox\pandoc@box{#1}%
  \Gscale@div\@tempa{\textheight}{\dimexpr\ht\pandoc@box+\dp\pandoc@box\relax}%
  \Gscale@div\@tempb{\linewidth}{\wd\pandoc@box}%
  \ifdim\@tempb\p@<\@tempa\p@\let\@tempa\@tempb\fi% select the smaller of both
  \ifdim\@tempa\p@<\p@\scalebox{\@tempa}{\usebox\pandoc@box}%
  \else\usebox{\pandoc@box}%
  \fi%
}
\def\fps@figure{htbp}
\makeatother
\NewDocumentCommand\citeproctext{}{}

\makeatletter
 \let\@cite@ofmt\@firstofone
 \def\@biblabel#1{}
 \def\@cite#1#2{{#1\if@tempswa , #2\fi}}
\makeatother
\newlength{\cslhangindent}
\newlength{\csllabelwidth}
\newenvironment{CSLReferences}[2] % #1 hanging-indent, #2 entry-spacing
 {\begin{list}{}{%
  \setlength{\itemindent}{0pt}
  \setlength{\leftmargin}{0pt}
  \setlength{\parsep}{0pt}
  \ifodd #1
   \setlength{\leftmargin}{\cslhangindent}
   \setlength{\itemindent}{-1\cslhangindent}
  \fi
  \setlength{\itemsep}{#2\baselineskip}}}
 {\end{list}}
\usepackage{calc}

\newcommand{\CSLLeftMargin}[1]{\parbox[t]{\csllabelwidth}{\strut#1\strut}}
\newcommand{\CSLRightInline}[1]{\parbox[t]{\linewidth - \csllabelwidth}{\strut#1\strut}}

\usepackage{ragged2e}
\usepackage{caption}
\usepackage{longtable}
\usepackage[labelformat = empty]{caption}
\usepackage{afterpage}
\usepackage{fontenc}
\usepackage[symbol]{footmisc}
\usepackage{mdframed}
\usepackage{float}
\definecolor{bleu}{HTML}{2200cc}
\newenvironment{shadebox}{\begin{figure*}\begin{mdframed}[backgroundcolor=black!10]} 
  {\end{mdframed}\end{figure*}} % this defines `shadebox` for shaded boxes that work with pandoc citations
\definecolor{bleu}{HTML}{2200cc}

\usepackage{booktabs}
\usepackage{longtable}
\usepackage{array}
\usepackage{multirow}
\usepackage{wrapfig}
\usepackage{float}
\usepackage{colortbl}
\usepackage{pdflscape}
\usepackage{tabu}
\usepackage{threeparttable}
\usepackage{threeparttablex}
\usepackage[normalem]{ulem}
\usepackage{makecell}
\usepackage{xcolor}
\usepackage{bookmark}
\IfFileExists{xurl.sty}{\usepackage{xurl}}{} % add URL line breaks if available
\makeatletter
\@ifundefined{xmpquote}{}{}
\makeatother
\hypersetup{
  colorlinks=true,
  linkcolor={bleu},
  filecolor={Maroon},
  citecolor={Blue},
  urlcolor={bleu},
  pdfcreator={LaTeX via pandoc}}

\author{}
\date{\vspace{-2.5em}}

\begin{document}

\raggedright
\LARGE

\textbf{Cultural invariance in music-evoked emotions}

\vspace{0.2in}

\justifying
\normalsize

Lidya Yurdum\(^{1,2,\ast}\), Jan Simson\(^{2,3,4,5}\), Roza G.
Kamiloğlu\(^{6,7}\), Disa A. Sauter\(^{1,8,\dagger}\), and Samuel A.
Mehr\(^{2,9,\dagger}\)

\small

\(^{1}\)Department of Psychology, University of Amsterdam, The
Netherlands\\
\(^{2}\)Child Study Center, Yale University, USA\\
\(^{3}\)University of Mannheim, Germany\\
\(^{4}\)Ludwig-Maximilians-Universität München, Germany\\
\(^{5}\)Munich Center for Machine Learning (MCML), Germany\\
\(^{6}\)Department of Experimental and Applied Psychology, Vrije
Universiteit Amsterdam, Amsterdam, The Netherlands\\
\(^{7}\)Institute Brain and Behaviour Amsterdam (iBBA), Vrije
Universiteit Amsterdam, Amsterdam, The Netherlands\\
\(^{8}\)Department of Experimental Psychology, University of Oxford,
United Kingdom\\
\(^{9}\)School of Psychology, University of Auckland, New Zealand\\

*Corresponding author. E-mail:
\href{mailto:l.yurdum@uva.nl}{\nolinkurl{l.yurdum@uva.nl}} (L.Y.)

\(\dagger\) These authors contributed equally; order was determined by
coin flip.

\vspace{0.2in}

\begin{shadebox}

\small

Worldwide, people produce and listen to music to induce, express, and
regulate their own and others' emotions\textsuperscript{1--7}. To what
extent culture influences the emotions evoked by music is, however,
contested. The documented global variation in musical
behaviours\textsuperscript{8--10} and emotion
concepts\textsuperscript{11--13} suggests substantial cross-cultural
differences in how listeners relate to music
emotionally\textsuperscript{14--16}, while shared physiology and
evolutionary history would predict broad global agreement on the
emotional information extracted from music\textsuperscript{5,17--21}.
Some evidence suggests shared processes\textsuperscript{22--28} but
experiments on the subject do not represent global diversity in either
music or participants\textsuperscript{29--32}. Here, we report a large
citizen-science experiment wherein listeners (\emph{N} = 124,394 in 189
countries; conducted in English) rated their emotional responses to
popular Western music and to vocal music produced in mostly small-scale
societies drawn from a globally representative corpus (i.e., songs
assumed to be unfamiliar in style). We found strong similarity in
listeners' emotional responses to music: Average agreement between pairs
of countries' listeners was high. Furthermore, across both popular and
unfamiliar types of music, emotional responses were shaped far more by
individual listeners and the specific songs than by cultural background:
Both between-person differences within a country and within-subject
differences across songs explained substantially more variance in
emotional responses than country-wise differences. Together, these
results indicate that a listener's cultural background has only a
marginal influence on the emotions they feel in response to music,
raising the possibility of universal structure in emotional responses to
music.

\end{shadebox}

\normalsize

\section*{Main text}\label{main-text}
\addcontentsline{toc}{section}{Main text}

Music likely exists in all cultures\textsuperscript{33} and has played a
central role in self-expression, ritual, and daily
life\textsuperscript{3,34,35} throughout our
history\textsuperscript{17,36}. A key feature of music is that it is
distinctly emotional: Listeners experience intense emotional reactions
when engaging with music\textsuperscript{1} and often listen to music to
alleviate negative feelings, enhance positive feelings, or set a
specific mood\textsuperscript{2--4}. Moreover, listeners readily
perceive emotions in music\textsuperscript{6,35} and can identify the
intended emotions behind compositions\textsuperscript{24}.

The centrality of emotions to music may help explain the ubiquity of
musical behaviours. Emotional responses are a key motivation for why
people seek out music: Listeners tend to prefer songs that induce
stronger emotions\textsuperscript{7} and the top reasons participants
report listening to music in their daily lives are also linked to
influencing and regulating emotions\textsuperscript{35}.

Emotional responses to music are likely the result of a complex set of
mechanisms\textsuperscript{37}, some of which are grounded in our shared
human physiology and evolutionary past\textsuperscript{38}. Our auditory
and autonomic nervous systems make certain acoustic features in music
(and other vocal channels\textsuperscript{39}) more likely to induce
certain emotions in listeners, regardless of cultural exposure. For
instance, sudden loudness or a fast-paced, steady beat are likely to
lead to arousal or excitement\textsuperscript{37,40}, while a slower
tempo and fewer changes in pitch are associated with
tenderness\textsuperscript{38} and relaxation\textsuperscript{41}. These
latter features are observed in lullabies across
cultures\textsuperscript{33,42} and their emotional associations develop
early in life: Western infants show physiological signs of relaxation
when hearing lullabies from unfamiliar cultures, despite having no prior
exposure to the musical traditions the lullabies originated
from\textsuperscript{43}.

The acoustic characteristics of music are predictive of the emotions
induced by music\textsuperscript{44,45}. As a result, listeners' reports
of the emotions induced by music may be expected to show similarity
across cultures. This prediction is supported by evidence from multiple
cultures\textsuperscript{22,23,46,47}. For instance, Western (European
and North American) and East Asian listeners (from China) agree on the
emotions induced by music from each other's
cultures\textsuperscript{28,48}.

Shared mechanisms do not preclude substantial cultural differences in
the emotions listeners experience when listening to music. Music is
evidently culturally situated and influenced by
experience\textsuperscript{16,49,50}: People tend to prefer familiar
music\textsuperscript{51}, are more sensitive to culturally-familiar
musical features\textsuperscript{47,52}, and preferences for some
musical characteristics differ cross-culturally\textsuperscript{53}.
More fundamentally, many aspects of musical behaviour are highly
culturally-specific, with structures and expressive functions that vary
across societies\textsuperscript{14,15}.

For instance, the meaning and emotions derived from Australian
Aboriginal navigational songs (\emph{songlines}) which provide a system
for navigation and invoke spiritual stories\textsuperscript{10}, are
difficult to comprehend for listeners unfamiliar with this cultural
context. Similarly, Western listeners may not be familiar with the
melodic patterns in Hindustani ragas that are culturally associated with
particular times of day and emotional states\textsuperscript{9}. The
emotions induced by music are thus influenced by culturally-learned
extra-musical and contextual information, such as familiarity with genre
conventions\textsuperscript{54} or knowledge of the original setting of
a piece\textsuperscript{55}.

Quantifying the extent of commonalities in the emotions induced by music
across cultures and across individuals can, however, be a challenge. In
practice, studies tend to focus on comparing two or three cultures at a
time, with small sets of musical examples\textsuperscript{29}. A lack of
diversity in both the listener and music samples\textsuperscript{30,31}
may skew estimates of cross-cultural agreement, in either a positive or
negative direction. In smaller studies, cross-cultural agreement may
reflect convergence within a narrow slice of global music and
populations rather than genuine cross-cultural agreement. Alternatively,
cases of cross-cultural disagreement may emerge as an exception that
proves the rule: an unusual case of disagreement as a result of the
specific stimuli and populations that are tested.

Citizen science and gamification present a set of tools that can help
address these concerns. In citizen-science research, participants are
motivated to voluntarily participate out of curiosity, interest,
enjoyment, or the desire to contribute to science, enabling the
collection of data from large numbers of volunteers\textsuperscript{56}.
Larger sample sizes enable the testing of large numbers of stimuli,
which increases the generalisability of findings\textsuperscript{32}. We
adopt this approach here, to ask whether cross-cultural agreement in
emotional responses to music emerges when both listeners and musical
stimuli are sampled to ensure high cultural diversity.

We recruited 124,394 participants living in 189 countries (Figure 1a),
who reported speaking a total of 218 languages other than English. They
participated in the study on
\href{https://themusiclab.org}{themusiclab.org}, a popular
citizen-science platform. Participants listened to brief excerpts of
popular Western music, with styles that were likely broadly familiar to
listeners (hereafter, popular music), and excerpts from the
\emph{Natural History of Song Discography}\textsuperscript{33},
featuring songs collected from mostly small-scale societies and sung in
non-English languages (hereafter, unfamiliar music; Figure 1b). After
each excerpt, participants indicated how the music made them feel by
choosing from a fixed set of emotion terms used in prior
research\textsuperscript{57}.

We asked how similar listeners' emotional responses to music were to one
another, and what explained links between musical content and emotional
responses. We first performed exploratory analyses using the popular
music dataset. Based on the results of these analyses, we pre-registered
the analyses we report for the unfamiliar music sample at
\url{https://osf.io/mw435} (see SI Text 1 for details). Readers can
explore the data visually and listen to the music featured in the study
at \url{https://lidyayurdum.shinyapps.io/evoke_app/}.

\begin{figure}

{\centering \includegraphics[width=0.8\linewidth]{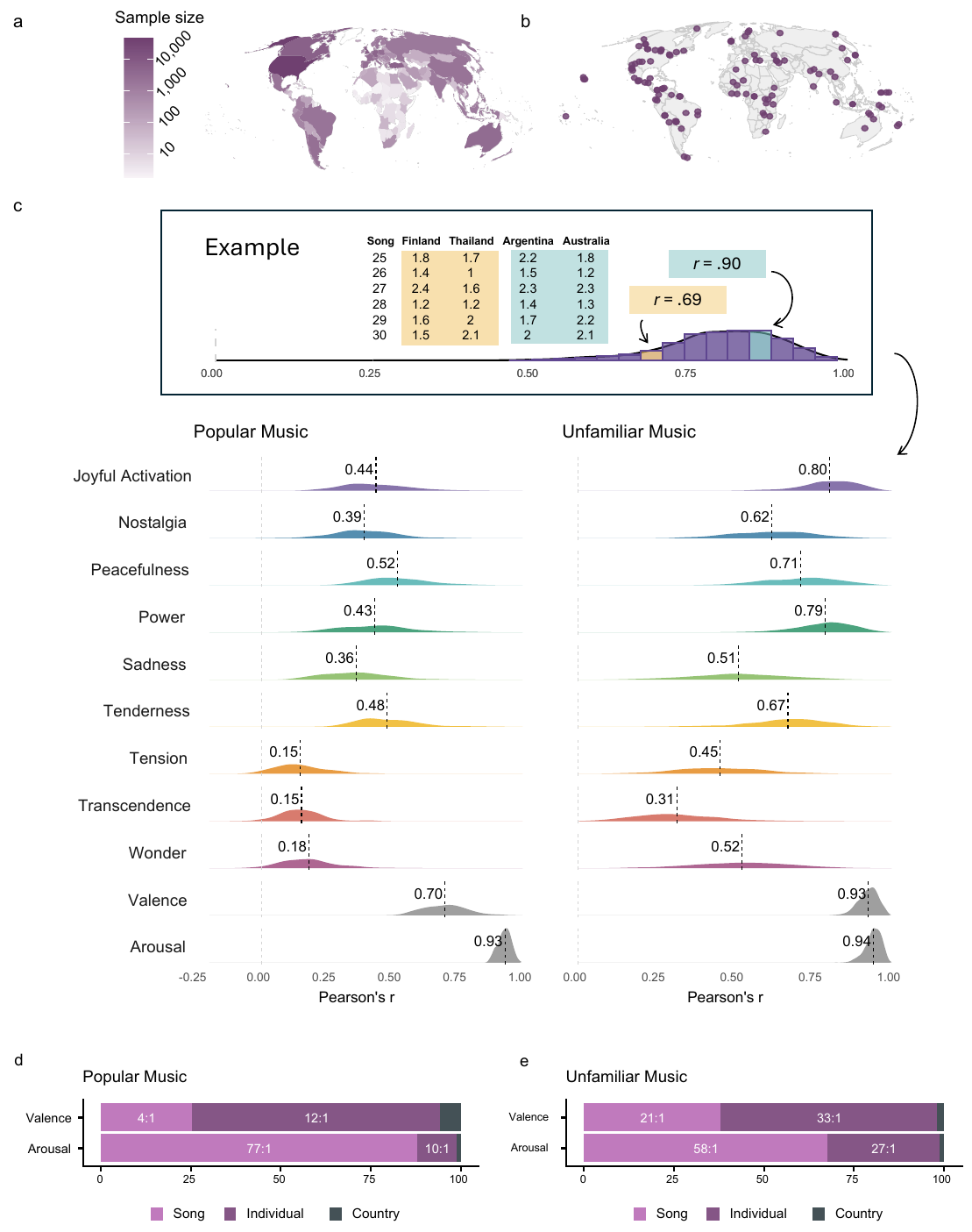} 

}

\caption{{\small \textbf{Figure 1. Relative invariance in emotional responses to music across countries.} (\textbf{a}) Geographic distribution of participants. (\textbf{b}) Locations of recordings studied, from the \textit{Natural History of Song Discography}, which comprised the unfamiliar musical stimuli. (\textbf{c}) Mean emotion ratings per song were significantly correlated between pairs of countries. To measure cross-country agreement, we first computed each country's mean rating on each emotion dimension for each song. For each pair of countries, we then correlated these song-wise means. This generated a correlation coefficient for each pair of countries, on each emotion dimension, plus arousal and valence (as demonstrated in the inset example). The histograms denote the distributions of Pearson's $r$ values for all pairwise country combinations, for popular (left side) and unfamiliar music (right side). Valence and arousal denote the two latent emotion dimensions produced by dimension reduction. Countries with fewer than 400 participants are excluded. (\textbf{d-e}) The percentage of explained variance attributable to the country the listener lives in, the song they heard, and the listener's own unique response pattern for (\textbf{d}) popular music and (\textbf{e}) unfamiliar music, from mixed-effects regressions predicting participants' emotional ratings with random intercepts for country, listener, and song. Valence and arousal denote the two latent emotion dimensions produced by dimension reduction (see Figure S5a-b for an equivalent visualisation of the GEMS-9 factors). The white overlay text indicates the ratios of variance explained by song to country or individual to country, respectively.}}\label{fig:unnamed-chunk-1}
\end{figure}

\section*{Results}\label{results}
\addcontentsline{toc}{section}{Results}

\subsection*{Emotional responses to music are highly consistent across
cultures}\label{emotional-responses-to-music-are-highly-consistent-across-cultures}
\addcontentsline{toc}{subsection}{Emotional responses to music are
highly consistent across cultures}

To facilitate cross-cultural analysis of emotional responses across many
listeners, songs, and emotion terms, we imputed missing data and reduced
the dimensionality of the emotion ratings, which yielded two primary
dimensions of variation corresponding to valence and arousal (see
Methods, SI Text 2, and Figures S1-2). The reduction of discrete emotion
ratings to these two factors is consistent with prior studies that
organise emotional responses to music along these
dimensions\textsuperscript{58--61}. Moreover, the two dimensions
correlated as expected with ratings of pleasantness and excitement for
the same songs from prior work\textsuperscript{62}, supporting our
interpretation of the dimensions as valence and arousal, respectively
(Figure S3).

First, we assessed agreement among groups of listeners by computing mean
scores for the valence and arousal factors for each song among listeners
from each country. We then computed all pairwise correlations among the
19 countries with sufficient data available for the analysis (\(n >\)
400; median \(n =\) 681, range = 434 - 14,357). The pairwise
correlations in popular music were high for both valence (median
\emph{r} = 0.70, IQR = 0.11) and arousal (median \emph{r} = 0.93, IQR =
0.03).

This pattern repeated across other analyses of interest. The
correlations remained high when using unimputed data (Figure 1c),
aggregated to produce the nine factors reported in previous
work\textsuperscript{57}, rather than our dimensionality reduction (SI
Text 3.1). The median correlation across all country pairings and all
nine emotion dimensions was \(r =\) 0.39 (range: \(r =\) 0.15 - 0.52,
IQR = 0.26 ; full reporting in Table S1). This latter analysis approach
likely represents a lower bound on the pairwise correlations; following
previous work\textsuperscript{28}, we also computed noise-corrected
correlation coefficients, which were higher (see Figure S4 and SI Text
4.1). We found comparable patterns of results when computing pairwise
correlations across language groups (Table S2).

Notably, cross-cultural similarity in emotion ratings was higher when
listeners evaluated unfamiliar vocal music, across imputed (median
\(r_{valence} =\) 0.93, median \(r_{arousal} =\) 0.94), unimputed
(median \(r =\) 0.62, range: \(r =\) 0.31 - 0.80, IQR = 0.20) and
noise-corrected correlations (median \(r =\) 0.77, range: \(r =\) 0.57 -
0.89, IQR = 0.12; full reporting in Table S1). This result suggests that
evaluating responses to familiar Western music, rather than more diverse
musical samples, can in fact underestimate cross-cultural similarities
in listeners. Unlike what might be expected from a standpoint of strong
cultural relativity, agreement was more substantial when listeners heard
music from cultures they had minimal exposure to, rather than songs that
featured familiar popular music characteristics.

A possible reason for the higher agreement among listeners when hearing
unfamiliar music could be the nature of the unfamiliar songs featured in
this study. The unfamiliar music corpus comprised songs used in four
specific behavioural contexts where singing occurs globally: soothing
infants, dancing, healing the sick, and expressing love for another
person\textsuperscript{33}. The strong agreement on the emotions induced
by these songs may reflect species-wide constraints on the forms of
songs used in specific behavioural contexts\textsuperscript{62}.
Functional constraints on the acoustic form of human songs have been
primarily argued in the contexts of soothing infants and coordinating
collective movement\textsuperscript{17,63,64}; and indeed, the emotional
ratings provided by listeners in the present study also most clearly
differentiated lullabies and dance songs (SI Text 5.1 and Figure S6).

\subsection*{Emotional responses to music vary more within cultures than
between
them}\label{emotional-responses-to-music-vary-more-within-cultures-than-between-them}
\addcontentsline{toc}{subsection}{Emotional responses to music vary more
within cultures than between them}

The above results suggest substantial agreement across cultures in the
emotions induced by music, and especially for unfamiliar traditional
music. What explains the variance that does exist in responses? We
addressed this question using a variance decomposition approach. We fit
mixed-effects regressions for each dimension, predicting participants'
emotion ratings with random intercepts for country, listener, and song,
so as to estimate the variance explained by each factor. We found that
the magnitudes of variability attributable to both individual
differences (i.e., how people's emotional responses vary in relation to
one another within a country) and of variability attributable to
differences in songs (i.e., how music itself varies emotionally) far
exceeded variability attributable to the listener's country of origin
(Figure 1d).

The proportions of variance explained by each factor are reported in
Table S3. For the recordings of popular music, the ratio of
within-country to between-country variability in emotion ratings was
12:1 for valence and 10:1 for arousal. The ratio of between-song to
between-country variability in emotion ratings was 4:1 for valence and
77:1 for arousal.

When analysing emotional responses to unfamiliar music (Figure 1e),
within-country variability again exceeded between-country variability
(33:1 for valence and 27:1 for arousal; Table S4). This implies that
cross-cultural similarity in emotion perception in music likely
generalises to \emph{all music}, rather than being limited to music from
familiar, popular-music styles. If anything, country-level information
about listeners explained less variability in listeners' emotional
responses when the music was from unfamiliar styles.

These results were not attributable to the specifications of the
pre-registered model: The same pattern of results was found when
analysing unimputed emotion ratings, aggregated to produce nine
factors\textsuperscript{57} (Figure S5a-b). They also were relatively
uninfluenced by listeners from countries with smaller samples; the main
findings repeated when limiting analyses to countries with at least 800
participants (instead of 400; SI Text 3.2). In contrast, when applying
the same model to a similarly structured dataset where cross-country
variability is known to be substantial (i.e., weather stations
distributed across the globe), a control analysis applied in previous
research\textsuperscript{33}, the results differed substantially, ruling
out the findings being an analytic artefact (Figure S5c and SI Text
3.3).

Last, to ensure that grouping listeners by country did not overestimate
group-level similarities, given the interconnectedness of modern
culture\textsuperscript{65,66}, we estimated variance ratios for the
nine factors from previous work\textsuperscript{57} using another proxy
for cultural similarity: the listeners' native language. The main
results replicated; the median ratio of within-language to
between-language variance for popular music across the factors was 26:1,
and 51:1 for unfamiliar music (full reporting in SI Text 3.4). These
findings suggest that listeners' cultural background has relatively
little influence on their emotional responses to music.

One limitation of these data is that the participants in this study all
completed the study in English --- a common limitation in cognitive
science research\textsuperscript{67}. If the participants in our study
were artificially homogeneous or particularly unique in how they respond
to the cues present in music, this could limit the generalisability of
findings. We think this is unlikely, as we did find a significant effect
of both country and language on the emotions reported by listeners,
indicating cultural heterogeneity in our sample. We also conducted a
secondary analysis to test the degree to which the results of the
English-language experiment reported here were in line with findings
from other experiments that were not conducted in English. Using an
external dataset\textsuperscript{68}, we found that our sample's emotion
ratings predicted perceptual judgements about the contexts in which the
same songs were used, made by native speakers of non-English languages,
and also by people living in three relatively isolated, smaller-scale
societies (see SI Text 5.2 and Figure S7). This correspondence suggests
that our sample's responses were not solely attributable to perceptual
judgements or biases unique to globalised, internet-connected
English-speakers.

\clearpage

\section*{Discussion}\label{discussion}
\addcontentsline{toc}{section}{Discussion}

In a large, diverse sample of listeners and a wide range of musical
examples, we found that the emotions listeners reported in response to
music were predictable from the song listeners heard, individual
variability, and country of origin --- but the magnitude of influence of
these variables differed considerably from one another. Country was
consistently the weakest predictor of emotional responses and
within-country differences were substantially larger than
between-country differences across two stimulus sets and all measures of
emotion. This result is surprising from a strong cultural-relativist
perspective, where the associations between musical features and
emotions would be learned primarily through enculturation and shaped by
local conventions.

The extent of agreement in our data instead speaks to species-wide
psychological and auditory mechanisms, around which individual
tendencies vary, resulting in larger within-country variation than
between. This is consistent with documented variability in musical
behaviours and features, which is concentrated within societies rather
than between them\textsuperscript{33,69}; but also with variability in
other aspects of human biology --- genetic variation within groups is
often much larger than variation between groups\textsuperscript{70}.
Similar results are also observable in non-musical behavioural and
psychological domains, such as parenting behaviours and
attitudes\textsuperscript{71}, personality and
values\textsuperscript{72}, the impressions people form of
others\textsuperscript{73}, and emotion regulation
strategies\textsuperscript{74}.

Given relatively large differences in sample size, we did not attempt to
model differences in the degree of cross-cultural agreement in responses
to the two types of music studied here (i.e., Western popular music and
unfamiliar traditional music). We note, however, that the raw levels of
listener agreement in response to culturally unfamiliar music are
substantially higher than those of Western popular music. This is not
necessarily inconsistent with prior work: Cross-cultural differences in
prior work are primarily observed when listeners from one culture are
familiar with the musical tradition\textsuperscript{22,27,75}. We
speculate that cultural differences in emotional responses may be
minimised when all listeners are equally unfamiliar with the music, as
in the present study. We also note that the traditional music studied
represents songs from behavioural contexts that reliably occur globally
(e.g., lullabies\textsuperscript{76}); it is possible that these songs
are functional in their given behavioural contexts precisely
\emph{because} they reliably evoke certain reactions. These results also
highlight a consideration for what musical characteristics are taken
into account when studying the relationship between music and emotions:
While responses are often contrasted across cultures or musical
traditions, the function a song serves is currently an under-studied
predictor of listener agreement. This may be particularly relevant when
a song is functional in social contexts that occur reliably globally, as
in the present study.

We found that the specific song listeners heard predicted listeners'
emotional reactions more strongly than the country they were from.
Acoustic features such as tempo, loudness and timbre can all shape the
emotions listeners experience when listening to
music\textsuperscript{77,78}, via mechanisms that may range from
low-level reflexes, to higher-level processes such as recognising the
emotion present in a singer's voice. The relatively stronger
contribution of acoustic information compared to culture-specific
knowledge is also evident in listeners' non-emotional judgements of
music. For example, our prior work found that acoustic information
(i.e., the particular song listeners heard) predicted how appropriate
listeners judged it for the functions of soothing infants, dancing,
expressing love, or healing illness. Familiarity with the song's origin
provided a statistically significant advantage in judgement accuracy,
but this boost associated with cultural learning was negligible compared
to the information present in the song itself\textsuperscript{68}.

Across both popular Western music and unfamiliar global music, we found
that the high-dimensional emotion space reduced to two dimensions,
broadly interpretable as arousal and valence. Arousal ratings were
primarily influenced by the song listeners heard while valence ratings
were more variable across individuals. This pattern is consistent with
prior work, and likely reflects the features these dimensions track:
arousal is more clearly linked to musical characteristics such as tempo
and amplitude, whereas valence tends to have less consistent acoustic
predictors\textsuperscript{e.g., 79,80}. Moreover, most of the musical
examples we studied included singing --- arousal, more so than valence,
tracks a range of acoustic parameters in non-musical vocal
prosody\textsuperscript{81--84}.

We do not, however, believe the present results imply that cultural
learning is unimportant in shaping emotional responses to music. In
other contexts, relatively small effects have helped to explain complex
psychological phenomena\textsuperscript{85}, including effects of
comparable size to those associated with country that we report here.
Furthermore, our study may not have created optimal conditions for
detecting cultural differences: we measured relatively immediate
responses to brief musical excerpts, without allowing participants to
personally select or engage deeply with the music. Intentional listening
and active engagement may provide greater opportunities for cultural
differences to emerge\textsuperscript{77}.

And while the songs studied were highly diverse, no single study can
fully represent the world's musical variation. The unfamiliar music
corpus, for example, features four song types that are thought to occur
universally (i.e., dance, lullaby, healing, love), but not other common
song types (e.g., children's play songs, mourning
songs\textsuperscript{86}). It also does not contain rarer examples of
culturally-specific uses of music, such as Mongolian long songs
(\emph{urtiin duu}), where single syllables can be drawn out such that a
song lasts up to three hours\textsuperscript{87}; or Tuvan throat
singing, where singers seek to harmonise with or mimic non-human natural
sounds\textsuperscript{88}.

We note several other limitations. First, country is an imperfect proxy
for cultural differences; and the countries with larger sample sizes
were also disproportionately likely to be Western (although the main
findings were also found when studying native languages rather than
countries of origin). Second, the study and all measurements were in
English, implying that participants were sufficiently literate in
English to complete the experiment. While this limitation is difficult
to avoid in large-scale cross-cultural work, its impact may be mitigated
by first reducing the dimensionality of responses to arousal and
valence: high-valence and high-arousal emotion terms tend to cluster
similarly across languages\textsuperscript{26} and emotional responses
to music have been characterised across 13 cultures in terms of arousal
and valence\textsuperscript{58}. Further, the unfamiliar music studied
was drawn from a pseudorepresentative global sample of music from
non-English-speaking, mostly small-scale societies. We also report
correspondence between the current (English-limited) data and prior
datasets collected in many non-English languages, including three
smaller-scale societies. Nonetheless, the English-language sampling of
participants may have reduced the extent of cultural differences
detected\textsuperscript{67}; future studies in other languages can
directly evaluate the scope of this limitation.

\bigskip

\section*{Methods}\label{methods}
\addcontentsline{toc}{section}{Methods}

\subsection*{Ethics approvals}\label{ethics-approvals}
\addcontentsline{toc}{subsection}{Ethics approvals}

All participants provided informed consent under protocols approved by
the Harvard University Committee on the Use of Human Subjects (protocol
IRB2017-1206) or the Yale University Human Research Protection Program
(protocol 2000033433). No compensation was provided.

\subsection*{Participants}\label{participants}
\addcontentsline{toc}{subsection}{Participants}

Participants were players of the \emph{What's Your Musical Style?} game
on \url{https://themusiclab.org}, a citizen science platform. The data
reported in this manuscript were collected between April 2019 and August
2026. Recruitment was largely by word-of-mouth, supplemented by social
media and mailing list postings.

We studied 124,394 participants whose reported age was between 16 and
100; who completed at least one trial of the music listening task; who
did not provide the same emotion intensity rating on every trial of the
task; and who did not have missing, duplicated, or internally
inconsistent country data (which occurs in online data collection due to
connectivity failures or other technical issues). These participants
provided a total of 3,472,453 individual emotion ratings, translating to
an average of 2,886 emotion ratings per song in the popular music corpus
and 22,678 per song in the unfamiliar music corpus.

Of these participants, 39,805 (43.6\% women, 52.6\% men, 3.8\% other,
0.02\% gender information missing or inconsistent) provided emotion
ratings for 276 songs from the popular music corpus (mean age = 24.41,
\emph{SD} = 9.4, median = 21). These listeners reported 165 different
native countries; the majority were from the United States (\emph{n} =
14,357; see Table S5 for demographics), Canada (\emph{n} = 1,923) or the
United Kingdom (\emph{n} = 1,821); 19 of the 165 countries reported by
listeners provided at least 400 participants (Table S6). A total of 170
different native languages were reported, with 13 languages represented
by at least 400 participants. English was the most commonly reported
first language (\emph{n} = 18,886). Spanish (\emph{n} = 3,063) and
German (\emph{n} = 1,700) were the next most widely spoken languages
(Table S7). Part way through data collection, an item regarding
bilingualism was introduced; of participants who provided this data,
54\% reported speaking a second language (\emph{n} = 21,454).

The 118 songs from the unfamiliar music corpus were rated by 123,836
participants (39.3\% women, 58.3\% men, 2.3\% other, 0.16\% missing or
inconsistent; mean age = 25.28, \emph{SD} = 9.42, median age = 22).
Listeners from 189 different countries provided data; the largest groups
were from the United States (\emph{n} = 44,844), Canada (\emph{n} =
6,295) or the United Kingdom (\emph{n} = 5,764). In total, 44 of the 189
countries reported by listeners provided at least 400 participants
(Table S8). A total of 219 different native languages were reported,
with 28 languages represented by more than 400 participants. English was
the most commonly reported first language (\emph{n} = 60,312), followed
by Spanish (\emph{n} = 11,813) and German (\emph{n} = 5,281; Table S9).
Of participants who provided data regarding bilingualism, 55\% reported
speaking a second language (\emph{n} = 28,056).

The global distribution of participants is visualised in Figure 1a.

\subsection*{Materials}\label{materials}
\addcontentsline{toc}{subsection}{Materials}

The game used stimuli from two musical corpora: a corpus of popular
music and a corpus of unfamiliar music. The former consisted of 276
(15-second) excerpts taken from songs that featured on the Billboard Hot
100 charts from 1951 to 2019. For each year, we selected the songs in
positions 1, 2, 19, and 20 on the year-end charts. The samples from the
years 1951-2013 were originally studied in prior
work\textsuperscript{89}. We supplemented this set with songs from the
2014-2019 charts, to keep the musical samples relatively up-to-date. The
samples contained either the chorus of the song or another recognisable
portion (full list of songs in Table S10).

The unfamiliar music corpus comprised 118 songs from the \emph{Natural
History of Song Discography}\textsuperscript{33}, recorded in 86 mostly
small-scale societies (Figure 1b). These songs are archival recordings
of vocal music, originally performed in the contexts of dancing,
soothing infants, healing illness, and expressing love for another
person (SI Text 5.1). Audio files and associated metadata are available
at \url{https://zenodo.org/records/7265514}.

\subsection*{Procedure}\label{procedure}
\addcontentsline{toc}{subsection}{Procedure}

The experiment was conducted in English. After providing informed
consent, participants were asked a set of demographic questions. These
included gender, age, country of residence, language, musical experience
and general level of education. Participants who had already provided
this information in a prior visit to the website were not necessarily
asked to re-report their responses.

Participants then listened to the song excerpts, drawn at random. The
game initially presented participants with 8 songs, randomly sampled
from the unfamiliar music corpus only; a further 8 songs randomly
sampled from the popular music corpus were added to the experiment at a
later date, resulting in 16 trials in total. After listening to each
song excerpt, participants first rated how much they liked the song
(n.b., we did not include or analyse those data for the work reported
here), then proceeded to the primary emotion rating task.

Participants were presented with three emotion labels randomly selected
from the 25-item version of the Geneva Emotional Music Scale
(GEMS)\textsuperscript{57}, one label at a time. Each item corresponds
to an emotion term (e.g., \emph{Joyful}, \emph{Tearful}, \emph{Dreamy}).
These items load on to the nine factors of the abbreviated version of
the scale, the GEMS-9\textsuperscript{57,90}. We made two changes to the
emotion labels: (1) the item \emph{Feeling of Transcendence} was changed
to \emph{Transcendent} in order to use consistent wording throughout the
study; and (2) the rating system was changed from five to four options,
to be consistent with other studies running on
\url{https://themusiclab.org}.

Participants rated how much the song made them feel each of the emotions
from 1 (Not at all) to 4 (Very much). These labels correspond to the
ones used in the GEMS\textsuperscript{57} with the exception of a
midpoint label (as there was no midpoint on the 4-point scale).
Participants were explicitly prompted to rate how the songs made them
\emph{feel}, rather than the emotion they thought the song expressed
(``That song made me feel {[}emotion{]}.''). At the end of the
procedure, the game presented participants with a song that they might
enjoy based on their responses, so as to increase participant engagement
and encourage word-of-mouth recruitment.

The study was built in \texttt{jsPsych}\textsuperscript{91}, a software
ecosystem for creating psychological experiments in JavaScript, and
hosted first via Pushkin\textsuperscript{92}, and later via
World-Wide-Lab (\url{https://worldwidelab.org/}).

\subsection*{Data imputation and dimension
reduction}\label{data-imputation-and-dimension-reduction}
\addcontentsline{toc}{subsection}{Data imputation and dimension
reduction}

We began by reducing the 25 GEMS items into a smaller set of underlying
dimensions, for two reasons. First, because the scale was originally
developed and validated mainly in Western samples, we preferred to allow
a unique cluster solution to emerge from our data, rather than assert
the nine factors found in prior work\textsuperscript{57}. Secondly,
because each listener rated only a subset of songs and provided
responses on a subset of the 25 possible emotion dimensions, our data
are sparse with data missing at random; reducing dimensionality helps
mitigate missingness by extracting stable underlying dimensions from the
available responses. We describe the dimension reduction procedure using
\texttt{softImpute}\textsuperscript{93} in SI Text 2.1.

Two data-driven components emerged. The loadings appeared to track
valence and arousal; we confirmed this statistically in two ways.

Firstly, we used the NRC VAD (Valence Arousal and Dominance)
lexicon\textsuperscript{94}, which contains human ratings of perceived
valence, arousal, and dominance for more than 55,000 English words. We
retrieved these ratings for each of the GEMS-25 emotion terms (e.g., the
term ``serene'', which is one of the GEMS-25 terms, scores high on
valence, but low on arousal), and correlated these values with the
loadings of each GEMS-25 dimension on the latent components. The
loadings of the two components retained in the popular music sample
significantly correlated with valence (\emph{r} = 0.76, \emph{p}
\textless{} .0001) and arousal (\emph{r} = 0.76, \emph{p} \textless{}
.0001), respectively. That is, the emotion terms that loaded most
strongly on the first component were also higher in valence, and the
emotion terms that loaded most strongly on the second component were
higher in arousal (Figure S1b-c). The second component also correlated
with dominance, though to a lesser extent (r = 0.61, \emph{p} = 0.001).
Similarly, two dimensions correlating with valence (\emph{r} = 0.72,
\emph{p} \textless{} .0001) and arousal (\emph{r} = 0.71, \emph{p}
\textless{} .0001; Figures S2b-c) emerged in the unfamiliar music sample
(with no significant correlations with dominance).

Secondly, we used data from a prior experiment using the same songs from
the unfamiliar music corpus, where participants rated pleasantness,
happiness, and excitement\textsuperscript{62}. All three terms are
characterised by positive valence, and pleasantness in particular is a
proxy for valence: Song-wise mean ratings of the three terms correlated
with the song-wise mean scores on the first component (pleasantness:
\emph{r} = 0.88, \emph{p} \textless{} .0001; happiness: \emph{r} = 0.63,
\emph{p} \textless{} .0001; excitement: \emph{r} = 0.70, \emph{p}
\textless{} .0001). The second component, by contrast, did not correlate
with pleasantness (pleasantness: \emph{r} = 0.01, \emph{p} = 0.95), but
it did correlate with excitement, which is a positively valenced high
arousal state (excitement: \emph{r} = 0.49, \emph{p} \textless{} .0001;
Figure S3) and to a lesser extent, happiness (happiness: \emph{r} =
0.41, \emph{p} \textless{} .0001).

The results of the dimensionality reduction were robust to sample size
specifications and also emerged when studying only countries with at
least 800 participants (Figure S8).

Given these convergent results, throughout this paper we describe the
first component as \emph{valence} and the second as \emph{arousal}. We
encourage readers to validate these dimensions by listening to the songs
used in the experiment while viewing their dimensional characterisation
in the Shiny app available at
\url{https://lidyayurdum.shinyapps.io/evoke_app/}.

\section*{End notes}\label{end-notes}
\addcontentsline{toc}{section}{End notes}

\subsection*{Data, code, and materials
availability}\label{data-code-and-materials-availability}
\addcontentsline{toc}{subsection}{Data, code, and materials
availability}

A fully reproducible manuscript; data, analysis code, and
visualizations; other materials; and code for the experiment are
available at \url{https://github.com/themusiclab/music-emotions}. Audio
files and metadata from the \emph{Natural History of Song Discography}
are available at \url{https://doi.org/10.5281/zenodo.7265514}. Readers
may participate in the experiment by visiting
\url{https://themusiclab.org/quizzes/jan} and view dynamic
visualizations at \url{https://lidyayurdum.shinyapps.io/evoke_app/}.

\subsection*{Acknowledgments}\label{acknowledgments}
\addcontentsline{toc}{subsection}{Acknowledgments}

We thank the participants who contributed to the study via
themusiclab.org. We are also grateful to the University of Amsterdam
Stats Store, and especially Dr.~Raoul Grasman, for guidance on dimension
reduction using \texttt{SoftImpute}; Dr.~Amy Belfi for collecting and
sharing the stimuli that comprised the popular music corpus; and members
of the Music Lab, especially Jingzuan Liu, Constance Bainbridge, Joyce
Seok, Anna Bergson, Brooke Milosh, Liam Crowley, and Mila Bertolo, for
reviewing stimuli metadata and providing feedback on the study design
and manuscript.

\subsection*{Author contributions}\label{author-contributions}
\addcontentsline{toc}{subsection}{Author contributions}

L.Y. wrote the manuscript, with contributions from D.S. and S.A.M.. L.Y.
conducted analyses and created visualisations, including the
accompanying Shiny App. D.S. and S.A.M. reviewed and edited the
manuscript, supervised the project, and acquired funding. R.G.K.
conducted code review. J.S. and S.A.M. conceived of the study. J.S.
programmed the experiment for \url{https://themusiclab.org}. All authors
edited and approved the resulting manuscript.

\subsection*{Competing Interests}\label{competing-interests}
\addcontentsline{toc}{subsection}{Competing Interests}

The authors declare no competing interests.

\subsection*{Funding}\label{funding}
\addcontentsline{toc}{subsection}{Funding}

This research was supported by the Harvard Data Science Initiative
(S.A.M.), the National Institutes of Health Director's Early
Independence Award DP5OD024566 (S.A.M.), the Royal Society of New
Zealand Te Apārangi Rutherford Discovery Fellowship RDF-UOA2103
(S.A.M.), and the Dutch Research Council Vici grant VI.C.221.052 (D.S.).

\subsection*{Artificial intelligence
disclosure}\label{artificial-intelligence-disclosure}
\addcontentsline{toc}{subsection}{Artificial intelligence disclosure}

We disclose the following uses of artificial intelligence: Author L.Y.
used the University of Amsterdam AI Chat (model \texttt{gpt-6.1-sol}) to
help polish selected sentences for clarity, suggest additional
literature to read, and generate and debug R code, primarily for
visualisations and Shiny App design. In all uses, authors verified
output manually.

\clearpage

\section*{References}\label{references}
\addcontentsline{toc}{section}{References}

\protect\phantomsection\label{refs}

\end{document}